\documentclass[epj]{svjour}
\usepackage{graphics}
\begin{document}
\title{Third order Lovelock black branes in the presence of a nonlinear electromagnetic field}
\author{S. H. Hendi\inst{1,2}\thanks{\emph{
hendi@shirazu.ac.ir}} \and S. Panahiyan\inst{1} \and H.
Mohammadpour\inst{1}
}                     
\institute{Physics Department and Biruni Observatory, Shiraz
University, Shiraz 71454, Iran \and Research Institute for
Astrophysics and Astronomy of Maragha (RIAAM), P.O. Box 55134-441,
Maragha, Iran}
\date{Received: date / Revised version: date}
%
\abstract{ We consider third order Lovelock gravity coupled to an
U(1) gauge field for which its Lagrangian is given by a power of
Maxwell invariant. In this paper, we present a class of horizon
flat rotating black branes and investigate their geometrical
properties and the effect of nonlinearity on the solutions. We use
some known formulas and methods to calculate thermodynamic and
conserved quantities. Finally, we check the satisfaction of the
first law of thermodynamics.
\PACS{
      {PACS-key}{discribing text of that key}   \and
      {PACS-key}{discribing text of that key}
     } 
} 
\maketitle
\section{Introduction}
Lovelock gravity is one of the higher derivative gravity theories,
natural generalization of Einstein's General Relativity,
introduced by David Lovelock \cite{Lovelock1,Lovelock2,Lovelock3}
in $1971$
\begin{equation}
I_{G}=\int d^{d}x\sqrt{-g}\sum\limits_{k=0}^{d/2}\alpha
_{k}\pounds _{k}, \label{LovelockAction}
\end{equation}
where ($\alpha _{k}$) is an arbitrary constant and ($\pounds
_{k}$) is the Euler density of a $2k$-dimensional manifold
\begin{equation}
\pounds _{k}=\delta _{\rho _{1}\sigma _{1}...\rho _{k}\sigma
_{k}}^{\mu _{1}\nu _{1}...\mu _{k}\nu _{k}}R_{\mu _{1}\nu
_{1}}^{\rho _{1}\sigma _{1}}...R_{\mu _{k}\nu _{k}}^{\rho
_{k}\sigma _{k}}.  \label{Eq(2)}
\end{equation}
In this equation, $\delta _{\rho _{1}\sigma _{1}...\rho _{k}\sigma
_{k}}^{\mu _{1}\nu _{1}...\mu _{k}\nu _{k}}$ is the generalized
totally anti-symmetric Kronecker delta and ($R_{\mu \nu }^{\rho
\sigma }$) is the Riemann tensor.

It has been shown that gravitational field equation, arising from
variation of the action \ref{LovelockAction}, only contains the
terms with at most second order derivatives of metric which
indicates the quantization of linearized Lovelock theory is free
of ghosts \cite{String1,String2}. Also, such theory may be used in
the context of AdS/CFT correspondence to investigate the effects
of including higher-curvature terms
\cite{AdSCFT1,AdSCFT2,AdSCFT3,AdSCFT4}. Another interesting
property of this theory comes from the fact that there is a
delicate relationship between Lovelock gravity and low energy
effective action of heterotic string theory in addition to the
Calabi--Yau compactifications of M-theory \cite{String1,String2}.

On the other hand, it has been shown that loop corrections of gravity \cite%
{Fradkin851,Fradkin852,Fradkin853,Fradkin854,Fradkin855} may be
lead to a nonlinear Born-Infeld type Lagrangian \cite{BI1,BI2} .
Hoffmann was the first one to attempt to couple the nonlinear
electrodynamics with gravity \cite{Hoffmann}. One of the current
interesting subjects in black hole physics is the investigation of
the effects of
nonlinear electrodynamic fields on the gravitational field \cite%
{BIpaper1,BIpaper2,BIpaper3,BIpaper4,BIpaper5,BIpaper6,BIpaper7,%
PMIpaper1,PMIpaper2,PMIpaper3,PMIpaper4,PMIpaper5,PMIpaper6,PMIpaper7%
,PMIpaper8,PMIpaper9,PMIpaper10,Ayon98,Oliveira94,Soleng}.
Considering nonlinear model of electrodynamic fields provide a
powerful laboratory to analyze the properties of the black hole
solutions.

In this paper, we obtain $d$-dimensional rotating black brane
solutions of the first four terms of Lovelock theory which are
cosmological constant, Einstein, Gauss-Bonnet and third order
Lovelock terms, in the presence of nonlinear electromagnetic field
and analyze their properties. We should note that third order
Lovelock term is an Euler density in six dimensions and in order
to have the contribution of all the above terms in the field
equation, the dimension of the spacetime should be equal to or
larger than seven. These solutions can be regarded as a
generalization of the Lovelock-Maxwell gravity
\cite{DehShah1,DehShah2}. We discuss some geometrical and
thermodynamic properties, like curvature, temperature and entropy.
We also make some comments on the effects of nonlinearity
parameter and Lovelock coefficients.

\section{Field Equations\label{Fiel}}

The action of third order Lovelock gravity in the presence of
power-Maxwell-invariant source may be written as
\begin{eqnarray}
\mathcal{I} &=&-\frac{1}{16\pi
}\int_{\mathcal{M}}d^{d}x\sqrt{-g}[R-2\Lambda
+\alpha _{2}\mathcal{L}_{2}+\alpha _{3}\mathcal{L}_{3}+  \nonumber \\
&&(\kappa \mathcal{F)}^{s}]+\mathcal{I}_{b},  \label{I}
\end{eqnarray}%
where $R$ is the Ricci scalar, $\Lambda $ is the cosmological constant, $%
\alpha _{i}$'s are Lovelock coefficients, $\mathcal{L}_{2}$ and $\mathcal{L}%
_{3}$ denote the Gauss-Bonnet Lagrangian and the third order
Lovelock term, given as
\begin{equation}
\mathcal{L}_{2}=R_{\mu \nu \gamma \delta }R^{\mu \nu \gamma \delta
}-4R_{\mu \nu }R^{\mu \nu }+R^{2},  \label{L2}
\end{equation}%
\begin{eqnarray}
\mathcal{L}_{3} &=&2R^{\mu \nu \sigma \kappa }R_{\sigma \kappa \rho \tau }R_{%
\phantom{\rho \tau }{\mu \nu }}^{\rho \tau }+8R_{\phantom{\mu
\nu}{\sigma
\rho}}^{\mu \nu }R_{\phantom {\sigma \kappa} {\nu \tau}}^{\sigma \kappa }R_{%
\phantom{\rho \tau}{ \mu \kappa}}^{\rho \tau }+  \nonumber \\
&&24R^{\mu \nu \sigma \kappa }R_{\sigma \kappa \nu \rho }R_{%
\phantom{\rho}{\mu}}^{\rho }+3RR^{\mu \nu \sigma \kappa }R_{\sigma
\kappa \mu \nu }-12RR_{\mu \nu }R^{\mu \nu }  \nonumber \\
&&+24R^{\mu \nu \sigma \kappa }R_{\sigma \mu }R_{\kappa \nu
}+16R^{\mu \nu }R_{\nu \sigma }R_{\phantom{\sigma}{\mu}}^{\sigma
+R^{3}}.  \label{L33}
\end{eqnarray}%
The last term in the first integral is the Lagrangian of
power-Maxwell-invariant theory, where $\kappa $ is an arbitrary
constant and $\mathcal{F}$ is the Maxwell invariant which is equal
to$\ F_{\mu \nu }F^{\mu \nu }$( $F_{\mu \nu }=\partial _{\mu
}A_{\nu }-\partial _{\nu }A_{\mu }$ is the electromagnetic tensor
field and $A_{\mu }$ is the vector potential) and the last term is
the boundary term which is chosen such that the variational
principle is well defined. Varying the action (\ref{I}) with
respect to $g_{\mu \nu }$ and $A_{\mu }$, one can obtain the
gravitational as well as the electromagnetic field equations
\begin{eqnarray}
&&G_{\mu \nu }^{(1)}+\Lambda g_{\mu \nu }+\alpha _{2}G_{\mu \nu
}^{(2)}+\alpha
_{3}G_{\mu \nu }^{(3)} =  \nonumber \\
&&-2\left[ \kappa sF_{\mu \rho }F_{\nu }^{\rho }\left( \kappa \mathcal{F}%
\right) ^{s-1}-\frac{1}{4}g_{\mu \nu }\left( \kappa \mathcal{F}\right) ^{s}%
\right] ,  \label{FieldEq}
\end{eqnarray}%
\begin{equation}
\partial _{\mu }\left( \sqrt{-g}\left( \kappa \mathcal{F}\right)
^{s-1}F^{\mu \nu }\right) =0,  \label{PMIEq}
\end{equation}%
where $G_{\mu \nu }^{(1)}=R_{\mu \nu }-\frac{R}{2}g_{\mu \nu }$ is
just the Einstein tensor, and $G_{\mu \nu }^{(2)}$ and $G_{\mu \nu
}^{(3)}$ are the second and third order Lovelock tensors obtained
as
\begin{eqnarray}
G_{\mu \nu }^{(2)} &=&2(R_{\mu \sigma \kappa \tau }R_{\nu }^{\phantom{\nu}%
\sigma \kappa \tau }-2R_{\mu \rho \nu \sigma }R^{\rho \sigma
}-2R_{\mu
\sigma }R_{\phantom{\sigma}\nu }^{\sigma }+RR_{\mu \nu })  \nonumber \\
&&-\frac{1}{2}\mathcal{L}_{2}g_{\mu \nu },  \label{G2}
\end{eqnarray}%
\begin{eqnarray}
G_{\mu \nu }^{(3)} &=&-3(4R^{\tau \rho \sigma \kappa }R_{\sigma
\kappa
\lambda \rho }R_{\phantom{\lambda }{\nu \tau \mu}}^{\lambda }-8R_{%
\phantom{\tau \rho}{\lambda \sigma}}^{\tau \rho
}R_{\phantom{\sigma
\kappa}{\tau \mu}}^{\sigma \kappa }R_{\phantom{\lambda }{\nu \rho \kappa}%
}^{\lambda }+  \nonumber \\
&&2R_{\nu }^{\phantom{\nu}{\tau \sigma \kappa}}R_{\sigma \kappa
\lambda \rho }R_{\phantom{\lambda \rho}{\tau \mu}}^{\lambda \rho
}-R^{\tau \rho \sigma
\kappa }R_{\sigma \kappa \tau \rho }R_{\nu \mu }+  \nonumber \\
&&8R_{\phantom{\tau}{\nu \sigma \rho}}^{\tau }R_{\phantom{\sigma
\kappa}{\tau \mu}}^{\sigma \kappa }R_{\phantom{\rho}\kappa }^{\rho }+8R_{%
\phantom {\sigma}{\nu \tau \kappa}}^{\sigma }R_{\phantom {\tau
\rho}{\sigma
\mu}}^{\tau \rho }R_{\phantom{\kappa}{\rho}}^{\kappa }  \nonumber \\
&&+4R_{\nu }^{\phantom{\nu}{\tau \sigma \kappa}}R_{\sigma \kappa
\mu \rho }R_{\phantom{\rho}{\tau}}^{\rho }-4R_{\nu
}^{\phantom{\nu}{\tau \sigma \kappa }}R_{\sigma \kappa \tau \rho
}R_{\phantom{\rho}{\mu}}^{\rho }+
\nonumber \\
&&4R^{\tau \rho \sigma \kappa }R_{\sigma \kappa \tau \mu }R_{\nu
\rho }+2RR_{\nu }^{\phantom{\nu}{\kappa \tau \rho}}R_{\tau \rho
\kappa \mu }+
\nonumber \\
&&8R_{\phantom{\tau}{\nu \mu \rho }}^{\tau
}R_{\phantom{\rho}{\sigma}}^{\rho
}R_{\phantom{\sigma}{\tau}}^{\sigma }+8R_{\phantom{\tau}{\nu \mu \rho }%
}^{\tau }R_{\phantom{\rho}{\sigma}}^{\rho }R_{\phantom{\sigma}{\tau}%
}^{\sigma }-  \nonumber \\
&&8R_{\phantom{\sigma}{\nu \tau \rho }}^{\sigma }R_{\phantom{\tau}{\sigma}%
}^{\tau }R_{\mu }^{\rho }-8R_{\phantom{\tau }{\sigma \mu}}^{\tau \rho }R_{%
\phantom{\sigma}{\tau }}^{\sigma }R_{\nu \rho
}-4RR_{\phantom{\tau}{\nu \mu
\rho }}^{\tau }R_{\phantom{\rho}\tau }^{\rho }  \nonumber \\
&&+4R^{\tau \rho }R_{\rho \tau }R_{\nu \mu
}-8R_{\phantom{\tau}{\nu}}^{\tau
}R_{\tau \rho }R_{\phantom{\rho}{\mu}}^{\rho }+4RR_{\nu \rho }R_{%
\phantom{\rho}{\mu }}^{\rho }  \nonumber \\
&&-R^{2}R_{\nu \mu })-\frac{1}{2}\mathcal{L}_{3}g_{\mu \nu }.
\label{G3}
\end{eqnarray}

The Lovelock supplementation, $\mathcal{I}_{b}$, is
Gibbons-Hawking-York boundary term which may be written as
\begin{equation}
\mathcal{I}_{b}=-\frac{1}{8\pi }\int_{\partial \mathcal{M}}d^{d-1}x\sqrt{%
-\gamma }\left[ K+\alpha _{2}L_{2b}+\alpha _{3}L_{3b}\right] ,
\label{Ib}
\end{equation}%
where%
\begin{equation}
L_{2b}=2\left( J-2\widehat{G}_{ab}^{(1)}K^{ab}\right) ,
\end{equation}
\begin{eqnarray}
L_{3b} &=&3(P-2\widehat{G}_{ab}^{(2)}K^{ab}-12\widehat{R}_{ab}J^{ab}+2%
\widehat{R}J-  \nonumber \\
&&4K\widehat{R}_{abcd}K^{ac}K^{bd}-8\widehat{R}_{abcd}K^{ac}K_{e}^{b}K^{ed}).
\end{eqnarray}%
In these equations $\gamma _{\mu \nu }$ and $K$ are, respectively,
induced
metric and the trace of extrinsic curvature of boundary, $\widehat{G}%
_{ab}^{(1)}$ and $\widehat{G}_{ab}^{(2)}$ denote the
$(d-1)$-dimensional
Einstein and second order Lovelock tensors (Eq. (\ref{G2})) of the metric $%
\gamma _{ab}$ while $J$ and $P$ are the traces of
\begin{eqnarray}
J_{ab} &=&\frac{1}{3}(2KK_{ac}K_{b}^{c}+K_{cd}K^{cd}K_{ab}-  \nonumber \\
&&2K_{ac}K^{cd}K_{db}-K^{2}K_{ab}),
\end{eqnarray}%
and
\begin{eqnarray}
P_{ab}
&=&\frac{1}{5}\{[K^{4}-6K^{2}K^{cd}K_{cd}+8KK_{cd}K_{e}^{d}K^{ec}-
\nonumber \\
&&6K_{cd}K^{de}K_{ef}K^{fc}+3(K_{cd}K^{cd})^{2}]K_{ab}-  \nonumber \\
&&(4K^{3}-12KK_{ed}K^{ed}+8K_{de}K_{f}^{e}K^{fd})K_{ac}K_{b}^{c}-
\nonumber
\\
&&24KK_{ac}K^{cd}K_{de}K_{b}^{e}+12(K^{2}-K_{ef}K^{ef})K_{ac}K^{cd}K_{db}
\nonumber \\
&&+24K_{ac}K^{cd}K_{de}K^{ef}K_{bf}\}.  \label{Pab}
\end{eqnarray}

\section{The $d$-dimensional Charged Rotating Black Branes\label{Sol}}

Now, we should consider a rotating spacetime and investigate its
properties. Since the rotation group and the number of independent
rotation parameters
in $d$-dimensions are, respectively, $SO(d-1)$ and the integer part of $%
(d-1)/2$, one of the rotating metrics with $k\leq \lbrack
(d-1)/2]$ rotation parameters and zero curvature boundary may be
written as
\begin{eqnarray}
ds^{2} &=&-f(r)\left( \Xi dt-{{\sum_{i=1}^{k}}}a_{i}d\phi _{i}\right) ^{2}+%
\frac{dr^{2}}{f(r)}+  \nonumber \\
&&\frac{r^{2}}{l^{4}}{{\sum_{i=1}^{k}}}\left( a_{i}dt-\Xi
l^{2}d\phi _{i}\right)
^{2}-\frac{r^{2}}{l^{2}}{\sum_{i<j}^{k}}(a_{i}d\phi
_{j}-a_{j}d\phi _{i})^{2}  \nonumber \\
&&+r^{2}{{\sum_{i=1}^{d-2-k}}}dx_{i}^{2},  \label{Metric}
\end{eqnarray}%
where $\Xi =\sqrt{1+\sum_{i}^{k}a_{i}^{2}/l^{2}}$ and the angular
coordinates are in the range $0\leq \phi _{i}\leq 2\pi $. Using
the suitable gauge potential ansatz
\begin{equation}
A_{\mu }=h(r)\left( \Xi \delta _{\mu }^{0}-a_{i}\delta _{\mu
}^{i}\right) \;{ (no\; sum\; on\; i),}  \label{Amu}
\end{equation}%
in Eq. (\ref{PMIEq}) leads to the following differential equations
\begin{equation}
Eq_{h}=(2s-1)rh^{\prime \prime }+(d-2)h^{\prime }=0,  \label{heq}
\end{equation}%
with the following solutions%
\begin{equation}
h(r)=\left\{
\begin{array}{cc}
-q\ln r & \;{for\; }s=\frac{d-1}{2} \\
-qr^{(2s-d+1)/(2s-1)} & \;{otherwise}
\end{array}%
\right. ,  \label{h(r)}
\end{equation}%
where prime and double prime denote first and second derivatives
with respect to $r$, respectively and $q$ is an integration
constant. It is worthwhile to note that, one can choose $s=d/4$ to
obtain inverse-square electric field. In this case the expression
of the electric field does not
depend on the dimension and its value coincides with the Reissner--Nordstr%
\"{o}m solution in four dimensions. In addition, in arbitrary
dimensions with $s=(d-1)/2$, the gauge potential is logarithmic
and it coincides with the charged BTZ solution. In order to have a
sensible asymptotic structure, the electromagnetic field should
vanish for large values of $r$. This condition leads to $s>1/2$.
In this paper we would like to investigate the solutions of
general case (i.e. $s>1/2$ and $s\neq (d-1)/2$). In addition, one
can find that $\mathcal{F}=-2\left( \frac{dh(r)}{dr}\right) ^{2}$
and the power of Maxwell invariant, $(\kappa \mathcal{F})^{s}$,
may be imaginary for positive $\kappa $, when $s$ is fractional.
Therefore, we set $\kappa =-1 $ to have real solutions without
loss of generality.

To find the metric function $f(r)$, we should consider the
components of Eq. (\ref{FieldEq}). At first, we consider the
static case ($a_{i}=0$). One may
show that the $tt$-component of Eq. (\ref{FieldEq}) may be written as%
\begin{eqnarray}
Eq_{tt} &=&\frac{r}{4}\left[ 6!\alpha _{3}(_{\phantom{n}{5}%
}^{d-2})f^{2}-4!\alpha _{2}(_{\phantom{n}{3}}^{d-2})r^{2}f+2\left(
d-2\right) r^{4}\right] f^{\prime }  \nonumber \\
&&+\left[ 6!(_{\phantom{n}{6}}^{d-2})\alpha _{3}f^{2}-4!\alpha _{2}(_{%
\phantom{n}{4}}^{d-2})r^{2}f+2(_{\phantom{n}{2}}^{d-2})r^{4}\right] \frac{f}{%
2}  \nonumber \\
&&+\Lambda r^{6}+\frac{(2s-1)\left( 2h^{\prime 2}\right)
^{s}r^{6}}{2} =0, \label{tt2}
\end{eqnarray}
where $(_{\phantom{n}{q}}^{\;p})=\frac{p!}{(p-q)!q!}$. Other
nonzero
components of Eq. (\ref{FieldEq}) are a combination of $\frac{dEq_{tt}}{dr}$%
, $Eq_{tt}$ and $Eq_{h}$. For example $x_{i}x_{i}$-component of Eq. (\ref%
{FieldEq}) can be written as
\[
Eq_{x_{i}x_{i}}=\frac{d}{dr}Eq_{tt}+\frac{(d-8)}{r}Eq_{tt}-\frac{%
sr^{5}\left( 2h^{\prime 2}\right) ^{s}}{h^{\prime }}Eq_{h}=0.
\]%
Since $Eq_{h}$ vanishes for obtained $h(r)$, it is sufficient to solve $%
Eq_{tt}=0$. For simplicity, we choose a special case
\begin{equation}
\alpha _{3}=\frac{(d-3)(d-4)\alpha _{2}^{2}}{3(d-5)(d-6)}.
\label{alpha3}
\end{equation}%
The solution of Eq. (\ref{tt2}) can be written as
\begin{equation}
f(r)=\frac{r^{2}}{(d-3)(d-4)\alpha _{2}}\left[ 1-g(r)^{1/3}\right]
, \label{f(r)}
\end{equation}%
where
\begin{eqnarray}
g(r) &=&1+\frac{6(d-3)(d-4)\alpha _{2}\Lambda }{\left( d-1\right) (d-2)}+%
\frac{3(d-3)(d-4)\alpha _{2}M}{r^{d-1}}  \nonumber \\
&&-\frac{3(d-3)(d-4)\alpha _{2}\left( 2s-1\right)
^{2}L(r)}{(d-2)\left(
2s-d+1\right) },  \label{gr} \\
L(r) &=&\left( \frac{2q^{2}\left( 2s-d+1\right) ^{2}}{%
(2s-1)^{2}r^{2(d-2)/(2s-1)}}\right) ^{s},  \nonumber
\end{eqnarray}%
and $M$\ is an integration constant. We should note that, in
addition to static components of Eq. (\ref{FieldEq}), Eq.
(\ref{f(r)}) satisfies all nonstatic components of Eq.
(\ref{FieldEq}). In addition, it is notable that this solution
reduces to the solution of Ref. \cite{DehShah2} for $s=1$, as it
should be.

\begin{figure}[tbp]
\resizebox{0.4\textwidth}{!}{  \includegraphics{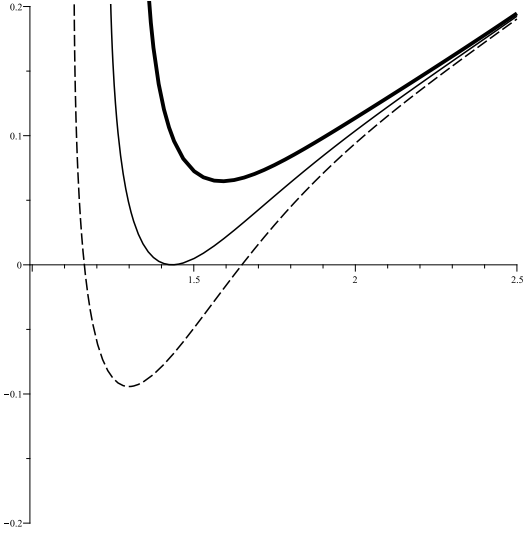}
} 
\caption{ $f(r)$ versus $r$ for $d=10$, $q=1$, $\Lambda=-1$, $\protect\alpha%
_{2}=0.1$, $s=2$, and $M=3.52<M_{\mathrm{ext}}$ (bold line), $M=4.52=M_{%
\mathrm{ext}}$ (solid line) and $M=5.52>M_{\mathrm{ext}}$ (dashed
line).} \label{Fig1}
\end{figure}

\begin{figure}[tbp]
\resizebox{0.4\textwidth}{!}{  \includegraphics{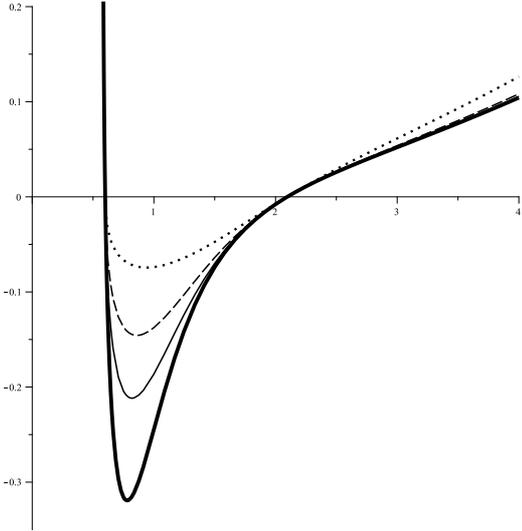}
} 
\caption{$f(r)$ versus $r$ for $d=7$, $q=1$, $\Lambda=-0.1$,
$M=1$, $s=2$, and $\protect\alpha_{2}=0.05$ (bold line),
$\protect\alpha_{2}=0.2$ (solid line), $\protect\alpha_{2}=0.5$
(dashed line) and $\protect\alpha_{2}=2$ (dotted line).}
\label{Fig2}
\end{figure}

\begin{figure}[tbp]
\resizebox{0.4\textwidth}{!}{  \includegraphics{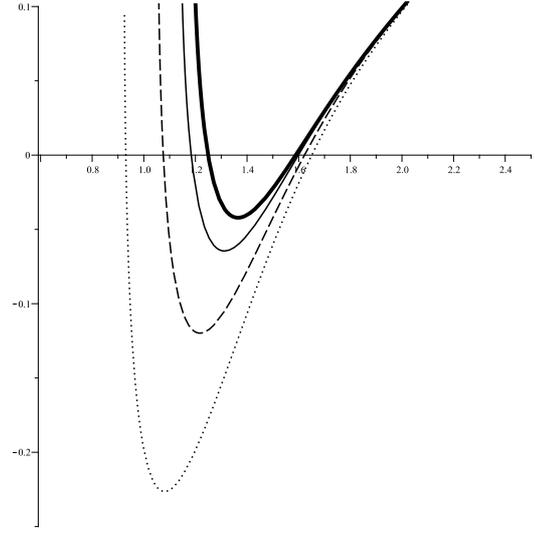}
} 
\caption {$f(r)$ versus $r$ for $d=10$, $q=1$, $\Lambda=-1$, $M=5$, $\protect%
\alpha_{2}=0.1$, and $s=2$ (bold line), $s=2.1$ (solid line),
$s=2.2$ (dashed line) and $s=2.3$ (dotted line).} \label{Fig3}
\end{figure}

Due to the fact that the power of $r$ in the denominator of $L(r)$
is bigger than unity for arbitrary $d>6$ and $s>1/2$, $L(r)$
vanishes for large values of $r$. Therefore in order to
investigate the asymptotic behavior of the solutions, one can
consider the vacuum solutions $(M=q=0)$
\begin{equation}
f(r)=\frac{r^{2}\left( 1-\left[ 1+\frac{6(d-3)(d-4)\alpha _{2}\Lambda }{%
\left( d-1\right) (d-2)}\right] ^{\frac{1}{3}}\right) }{(d-3)(d-4)\alpha _{2}%
}.  \label{Fg0}
\end{equation}%
Considering the former equation, it is easy to find that the
asymptotic
behavior of the solutions is AdS for negative $\Lambda $. One can put $%
\Lambda =-(d-1)(d-2)/2l^{2}$ in Eq. (\ref{Fg0}) to find the
effective cosmological constant of the solutions
\begin{equation}
\Lambda _{\mathrm{eff}}=-\frac{(d-1)(d-2)\left( 1-\left[ 1-\frac{%
3(d-3)(d-4)\alpha _{2}}{l^{2}}\right] ^{\frac{1}{3}}\right) }{%
2(d-3)(d-4)\alpha _{2}}.  \label{leff}
\end{equation}%
Now, we look for the essential singularity(ies). It is easy to
show that the Kretschmann scalar of the metric (\ref{Metric}) is
\begin{equation}
R_{\alpha \beta \gamma \delta }R^{\alpha \beta \gamma \delta
}=f^{\prime \prime 2}+2(d-2)\frac{f^{\prime
2}}{r^{2}}+2(d-2)(d-3)\frac{f^{2}}{r^{4}}. \label{RR}
\end{equation}%
After some algebraic manipulation, one can show that the Kretschmann scalar (%
\ref{RR}) with metric function (\ref{f(r)}) diverges at $r=0$ and
therefore there is an essential timelike singularity located at
$r=0$. In order to
investigate the existence of horizon(s), one should find the root(s) of $%
g^{rr}=f(r)=0$. In addition, we can use the fact that the
temperature of the extreme black brane is zero. In the case of
extreme black brane solution, the only real root of $f(r)$ is its
minimum. So, it is easy to find that the mass parameter for this
case becomes
\begin{equation}
m_{\mathrm{ext}}=\frac{4\Lambda s\left( \frac{-2\Lambda }{\left( \frac{%
2(2s-d+1)^{2}q_{\mathrm{ext}}^{2}}{(2s-1)^{2}}\right) ^{s}(2s-1)}\right) ^{%
\frac{(d-1)(2s-1)}{\left[ 2s(d-2)\right] }}}{(2s-d+1)(d-1)}.
\label{Mext}
\end{equation}%
Finally, we should note that obtained solutions may be interpreted
as black branes with inner and outer event horizons provided
$m>m_{\mathrm{ext}}$, extreme black brane for $m=m_{\mathrm{ext}}$
and naked singularity otherwise (see Fig. \ref{Fig1} for more
details). In addition, it is desirable to investigate the effects
of the Lovelock parameter, $\alpha _{2}$ and the nonlinearity
parameter, $s$. For example, figures \ref{Fig2} and \ref{Fig3}
show that both $\alpha _{2}$ and $s$ affect on the minimum value
of $f(r)$. Also, these figures confirm that in spite of $\alpha
_{2}$, the nonlinearity parameter, $s$ has a considerable role on
the values of inner and outer horizons.

\section{Thermodynamic and Conserved Quantities and the first law\label%
{Thermo}}

Now, we are in a position to calculate the thermodynamic and
conserved quantities of the solutions. The Hawking temperature and
angular velocities of the event horizon of the black branes may be
calculated by requiring the absence of conical singularity at the
horizon in the Euclidean sector of the
solutions. Considering the analytic continuation of the metric by setting $%
t\longrightarrow i\tau $ and $a_{i}\longrightarrow ia_{i}$, and
regularity at event horizon ($r=r_{+}$) help us to obtain the
Hawking temperature and the angular velocities of the black branes
\begin{eqnarray}
{T}_{+} &=&\mathcal{\beta }_{+}^{-1}{=}\frac{f^{\prime
}(r_{+})}{4\pi \Xi }=
\nonumber \\
&&{{\frac{r_{+}\left[ (1-2s)\left( \frac{2q^{2}\left( 2s-d+1\right) ^{2}}{%
(2s-1)^{2}r_{+}^{2(d-2)/(2s-1)}}\right) ^{s}-2{\Lambda }\right]
}{4(d-2)\pi \Xi }}},  \label{Tem}
\end{eqnarray}%
\begin{equation}
\Omega _{i}=\frac{a_{i}}{\Xi l^{2}}.  \label{Om}
\end{equation}%
Using the fact that the temperature of the extreme black branes
vanishes, one can write the horizon radius of the extreme black
brane
\begin{equation}
r_{\mathrm{ext}}^{2s(d-2)/(2s-1)}=\frac{(\frac{1}{2}-s)}{{\Lambda
}}\left( \frac{2q_{\mathrm{ext}}^{2}\left( 2s-d+1\right)
^{2}}{(2s-1)^{2}}\right) ^{s}.  \label{rext}
\end{equation}

In order to obtain the electric charge of the solution, we use the
generalized Maxwell equation (\ref{PMIEq}). Calculating the flux
of the
electric field at infinity leads to the electric charge per unit volume $%
V_{d-2}$
\begin{equation}
Q=\frac{2^{s}\Xi s}{8\pi }\left( \frac{\left( 2s-d+1\right)
q}{2s-1}\right) ^{2s-1}.  \label{Charge}
\end{equation}
The electric potential $U$, measured at infinity with respect to
the horizon, is calculated by the following relation
\begin{equation}
U=A_{\mu }\chi ^{\mu }\left\vert _{r\rightarrow \infty }-A_{\mu
}\chi ^{\mu }\right\vert _{r=r_{+}},  \label{U}
\end{equation}
where $\chi =\partial _{t}+\sum_{i}^{k}\Omega _{i}\partial _{\phi
_{i}}$ is the Killing vector. The electric potential will be
\begin{equation}
U=\frac{q}{\Xi }r_{+}^{(2s-d+1)/(2s-1)}.  \label{Pot}
\end{equation}

Now, we should calculate the total mass density of the black
branes. We may obtain finite mass through the use of the
counterterm method \cite{Mal}. One
can show that for obtained solutions with flat boundary, $\widehat{R}%
_{abcd}(\gamma )=0$, the finite action is \cite{DehShah2,DBS}
\begin{equation}
\mathcal{I}_{finite}=\mathcal{I}+\frac{1}{8\pi }\int_{\partial \mathcal{M}%
}d^{d-1}x\sqrt{-\gamma }\left( \frac{d-2}{l_{eff}}\right) .
\label{Ifinite}
\end{equation}%
It is notable that that this counterterm has exactly the same form
as the counterterm in Einstein gravity for a spacetime with zero
curvature boundary in which $l$ is replaced by $l_{eff}$. After
some algebraic manipulation, one can show that $l_{eff}$\ is given
by
\begin{eqnarray}
l_{eff} &=&\frac{15\sqrt{(1-\lambda )(d-3)(d-4)\alpha _{2}}}{9\left( 1+\frac{%
(d-3)(d-4)\alpha _{2}}{l^{2}}\right) -\left( 2+\lambda \right)
^{2}},
\label{Leff} \\
\lambda &=&\left( 1-\frac{3(d-3)(d-4)\alpha _{2}}{l^{2}}\right)
^{1/3}, \nonumber
\end{eqnarray}%
which reduces to $l$\ as $\alpha $\ goes to zero. Using Eqs.
(\ref{I}) and (\ref{Ifinite}) with Eq. (\ref{Leff}), the finite
action per unit volume $V_{d-2}$ can be obtained as
\begin{equation}
I=\frac{\mathcal{\beta }_{+}r_{+}^{d-1}\left[ {{\frac{{\Lambda }}{(d-1)}}}+%
\frac{2^{s-1}q^{2s}(2s-2d+3)}{r_{+}^{2s(d-2)/(2s-1)}\left(
\frac{\left( 2s-d+1\right) }{2s-1}\right) ^{1-2s}}\right] }{8\pi
(d-2)}. \label{finiteAct}
\end{equation}%
Having the total finite action, we can use the Brown-York method
of quasilocal definition \cite{Brown} to introduce the following
divergence-free stress-energy tensor
\begin{eqnarray}
T^{ab} &=&\frac{1}{8\pi }\{(K^{ab}-K\gamma ^{ab})+2\alpha
_{2}(3J^{ab}-J\gamma ^{ab})  \nonumber \\
&&\ +3\alpha _{3}(5P^{ab}-P\gamma ^{ab})+\frac{d-2}{l_{eff}}\gamma
^{ab}\ \}. \label{Tab}
\end{eqnarray}%
It is easy to find that $\partial /\partial t$ and $\partial
/\partial \phi ^{i}$ are the Killing vectors of the metric
(\ref{Metric}) and therefore the associated conserved quantities
are the mass and angular momentum, which are
\begin{eqnarray}
M &=&\frac{V_{d-2}}{16\pi }m\left[ (d-1)\Xi ^{2}-1\right] ,  \label{Mass} \\
J_{i} &=&\frac{V_{d-2}(d-1)}{16\pi }\Xi ma_{i}.  \label{Angmom}
\end{eqnarray}%
We should note that for static solutions ($a_{i}=0$ and therefore
$\Xi =1$), the angular momentum vanishes, and thus the $a_{i}$'s
are the rotational parameters of the spacetime.

In order to complete this section, we should calculate the entropy
of the black branes. Since the area law is applicable to black
holes (branes) in Einstein gravity
\cite{Arealaw1,Arealaw2,Arealaw3,Arealaw4,Arealaw5,Arealaw6,Arealaw7},
we calculate the entropy through the use of Gibbs--Duhem relation
\begin{equation}
\mathcal{S}=\frac{1}{T}\left( M-QU-{{\sum_{i=1}^{k}}}\Omega
_{i}J_{i}\right) -I.  \label{GibsDuh}
\end{equation}
It is strightforward to calculate the entropy per unit volume
$V_{d-2}$by use of Eq. (\ref{GibsDuh}) with obtained thermodynamic
and conserved quantities
\begin{equation}
\mathcal{S}=\frac{\Xi }{4}r_{+}^{d-2},  \label{Entropy}
\end{equation}
which confirms the entropy obeys the area law for our case (flat
horizon).

Now, we are in a position to check the first law of
thermodynamics. It has been found that the first law may be
derived from Smarr(-type) formula for the total energy of black
holes \cite{Smarr}. In order to obtain Smarr-type formula, one can
calculate the mass density as a function of the extensive
quantities $\mathcal{S}$, $\mathbf{J}$, and $Q$. Using Eqs. (\ref{Charge}), (%
\ref{Mass} ), (\ref{Angmom}), (\ref{Entropy}) and the fact that
$r_{+}$ is
the largest root of $f(r)$, we obtain%
\begin{equation}
M(\mathcal{S},\mathbf{J},Q)=\frac{\left[ (d-1)Z-1\right] J}{(d-1)l\sqrt{%
Z(Z-1)}},  \label{Smarr}
\end{equation}%
where $J=\left\vert \mathbf{J}\right\vert
=\sqrt{\sum_{i}^{k}J_{i}^{2}}$ and
$Z=\Xi ^{2}$ is the positive real root of the following equation%
\begin{eqnarray}
&&\frac{\left[ \frac{2^{s-1}(d-1)\left( 2s-1\right) ^{2}}{\left(
2s-d+1\right) }\left( \frac{2^{1-s}}{s}\frac{\pi Q}{\mathcal{S}}\right) ^{%
\frac{2s}{2s-1}}+\Lambda \right] l\left( 4\mathcal{S}\right) ^{\frac{d-1}{d-2%
}}}{8\pi (d-2)J}-  \nonumber \\
&&\sqrt{\frac{Z^{\frac{1}{d-2}}}{Z-1}} =0.  \label{Zeq}
\end{eqnarray}%
Now, we can regard the mass $M(\mathcal{S},\mathbf{J},Q)$ as a
function of extensive parameters and obtain the intensive
parameters conjugate to them in the following manner
\begin{equation}
T=\left( \frac{\partial M}{\partial \mathcal{S}}\right) _{J,Q},\ \
\Omega _{i}=\left( \frac{\partial M}{\partial J_{i}}\right)
_{\mathcal{S},Q},\ \ U=\left( \frac{\partial M}{\partial Q}\right)
_{\mathcal{S},J}. \label{Dsmar}
\end{equation}%
Using the chain rule, it is straightforward to show that the
intensive
quantities calculated by Eq. (\ref{Dsmar}) coincide with Eqs. (\ref{Tem}), (%
\ref{Om}) and (\ref{Pot}), and hence we can deduce that our black
brane solutions satisfy the first law of thermodynamics
\begin{equation}
dM=Td\mathcal{S}+{{{\sum_{i=1}^{k}}}}\Omega _{i}dJ_{i}+UdQ.
\label{1stLaw}
\end{equation}
\section{CLOSING REMARKS}

In this paper, we regarded the first four terms of Lovelock
gravity with a source of nonlinear electromagnetic field. We found
a class of rotating solutions which may be interpreted as black
branes with inner and outer horizons, extreme black branes or
naked singularity.

We used the analytic continuation of the metric to calculate the
Hawking temperature and angular velocities of the black branes. We
then considered the Gauss's law to obtain the finite electric
charge of the black branes. In addition, we calculated the finite
action, mass and angular momentum by using the counterterm method.
Since the area law can be applied to black holes (branes) in
Einstein gravity, we calculated the entropy through the use of
Gibbs--Duhem relation and found that the entropy obeys the area
law.

The Smarr-type formula for the mass was also obtained as a
function of extensive quantities to compute temperature, angular
velocities and electric potential. It was then confirmed that the
thermodynamic and conserved quantities satisfy the first law of
thermodynamics.

Finally, it is also desirable to study the causal structure, the
ratio of shear viscosity to entropy density, thermodynamic
stability and dynamical properties of the black brane solutions
derived here. Generalization of these solutions to various
(nontrivial) horizon topologies, remain to be carried out in the
future. In addition, It would be also interesting to investigate
the effect of curvature-cubed Lagrangian
\cite{quasitop11,quasitop12} and quartic quasitopological gravity
\cite{quasitop2} with a nonlinear source. We hope to address these
issues in the future.

\begin{acknowledgement}
\textbf{Acknowledgement:} We thank the referee for constructive
comments. This work has been supported financially by Research
Institute for Astronomy \& Astrophysics of Maragha (RIAAM).
\end{acknowledgement}

\end{document}